\documentclass[fleqn,usenatbib]{mnras}

\usepackage{newtxtext,newtxmath}

\usepackage[T1]{fontenc}

\DeclareRobustCommand{\VAN}[3]{#2}
\let\VANthebibliography\thebibliography
\def\thebibliography{\DeclareRobustCommand{\VAN}[3]{##3}\VANthebibliography}

\usepackage{graphicx}	
\usepackage{amsmath}	

\usepackage{graphicx}
 \usepackage{setspace}
\usepackage{textcomp}     
\usepackage{url}
\usepackage[]{amsmath}
\usepackage{amsmath}
\usepackage{nccmath}
\usepackage{lineno}

\newcommand{\av}{{2020 AV$_2$ }}
\newcommand{\avns}{{2020 AV$_2$}}
\newcommand{\an}{{\textquotesingle Ayl\'{o}\textquotesingle chaxnim }}
\newcommand{\anns}{{\textquotesingle Ayl\'{o}\textquotesingle chaxnim}}

\title[A kilometre sized asteroid inside Venus' orbit]{The discovery and characterization of a kilometre sized asteroid inside the orbit of Venus}

\author[B. T. Bolin et al.]{
Bryce T. Bolin,$^{1,2}$\thanks{E-mail: bbolin@caltech.edu (BTB)}
T. Ahumada,$^{3}$
P. van Dokkum,$^{4}$
C. Fremling,$^{1}$
M. Granvik,$^{5,6}$
\newauthor
K. K. Hardegree-Ullman,$^{7}$\thanks{Visiting astronomer, Cerro Tololo Inter-American Observatory at NSF's NOIRLab, which is managed by the Association of Universities for Research in Astronomy (AURA) under a cooperative agreement with the National Science Foundation.}
Y. Harikane,$^{8,9}$
J. N. Purdum,$^{10}$
E. Serabyn,$^{11}$
\newauthor
J. Southworth$^{12}$
and C. Zhai$^{11}$
\\
$^{1}$Division of Physics, Mathematics and Astronomy, California Institute of Technology, Pasadena, CA 91125, USA,\\
$^{2}$Infrared Processing and Analysis Center, California Institute of Technology, Pasadena, CA 91125, USA,\\
$^{3}$Department of Astronomy, University of Maryland, College Park, MD 20740, USA, 
$^{4}$Department of Astronomy, Yale University, New Haven, CT 06511, USA,\\
$^{5}$Department of Physics, University of Helsinki, 00560, Finland,
$^{6}$Asteroid Engineering Lab, Lule\r{a} University of Technology, Kiruna, 981 28 Sweden,\\
$^{7}$Steward Observatory, University of Arizona, Tucson, AZ 85721, USA,
$^{8}$Institute for Cosmic Ray Research, The University of Tokyo Kashiwa, Chiba 277-8582, Japan,\\
$^{9}$National Astronomical Observatory of Japan, Tokyo, 181-8588, Japan,
$^{10}$Caltech Optical Observatory, California Institute of Technology, Pasadena, CA 91125, USA,\\
$^{11}$Jet Propulsion Laboratory, California Institute of Technology, Pasadena, CA 91109, USA,\\
$^{12}$Astrophysical Research Institute, Liverpool John Moores University, Liverpool, L2 2QP, UK
}
\date{Accepted XXX. Received YYY; in original form ZZZ}

\pubyear{2022}

\begin{document}
\label{firstpage}
\pagerange{\pageref{firstpage}--\pageref{lastpage}}
\maketitle

\begin{abstract}
Near-Earth asteroid population models predict the existence of bodies located inside the orbit of Venus. Despite searches up to the end of 2019, none had been found. We report discovery and follow-up observations of (594913) \anns, an asteroid with an orbit entirely interior to Venus. (594913) \an has an aphelion distance of $\sim$0.65 au, is $\sim$2~km in diameter and is red in colour. The detection of such a large asteroid inside the orbit of Venus is surprising given their rarity according to near-Earth asteroid population models. As the first officially numbered and named asteroid located entirely within the orbit of Venus, we propose that the class of interior to Venus asteroids be referred to as \an asteroids.
\end{abstract}




\section{Introduction}

Almost all of the $\sim$1 million known asteroids are located on orbits exterior to Earth's orbit compared to just a fraction of a percent that have orbits located entirely inside Earth's orbit \citep[][]{Binzel2015}.  Dynamical models predict that a small fraction of the near-Earth asteroid (NEA) population \citep[][]{Bottke2002a,Granvik2018} consists of Atira asteroids located between the orbit of the Earth and Venus, and inner-Venus asteroids (IVAs) located entirely within the orbit of Venus.  However, IVAs have not been observed despite previous searches for objects interior to the orbit of the Earth \citep[][]{Whitely1998, Zavodny2008,Ip2021supp}. This is in part due to the difficulty of surveying this region of the Solar System within a small angular distance of the Sun with ground-based telescopes \citep[][]{Masi2003}. This postulated IVA population has been provisionally
 referred to as Vatiras\footnote{"Provisional because it will be abandoned once the first discovered member of this class will be named." \citep[][]{Greenstreet2012vat}.}, by analogy with the Atiras \citep[][]{Greenstreet2012vat}.
 
 The Zwicky Transient Facility (ZTF) mounted on the Palomar Observatory Samuel Oschin Schmidt Telescope is an all-sky survey designed to detect transients in the northern hemisphere \citep[][]{Bellm2019, Ip2021supp}. A portion of the time for the ZTF survey is designed to observe portions of the sky as close as possible to the Sun during evening and morning twilight called the ``Twilight Survey'' \citep[][]{Ip2021supp}. A preliminary version of the Twilight Survey ran in late 2018 and the first half of 2019 \citep[][]{Ye2020rr}. An expanded version of the Twilight Survey was executed between 2019 September 20 and 2020 January 30, observing during astronomical twilight on each clear night. 
 
\section{Observations}

The Twilight Survey is scheduled within the framework of the ZTF scheduler \citep[][]{Bellm2019b}. The Twilight Survey alternates between evening and morning twilight providing a total of 90 observing sessions, on 47 mornings and on 43 evenings between 2019 September 20 and 2020 January 30. The scheduler produces a nearly connecting 10 field pattern that covers 470 square degrees of sky during each observing session at elevations down to $\sim$20 degrees. Each Twilight Survey session lasts for 20-25 minutes. Each field in the pattern is imaged four times with 30\,s exposures in $r$-band (wavelength $\sim$680 nm) \citep[][]{Bellm2019b}. The time between subsequent exposures per single Twilight Survey fields is $\sim$5 minutes. The time spacing of the Twilight Survey cadence enables the detection of objects moving in the range between $\sim$8 arcseconds per hour and $>$1500 arcseconds per hour \citep[][]{Masci2019, Duev2019}. 

The sensitivity of the Twilight Survey images is brighter than the nominal ZTF limiting $V$-band (wavelength $\sim$550~nm) magnitude of $\sim$21 \citep[][]{Bellm2019}. The brighter sensitivity is due to the higher airmass of the observations combined with the higher sky background during astronomical twilight, resulting in a limiting magnitude close to $V\sim$20 \citep[Fig.~A1,][]{Ip2021supp}. A total of $\sim$40,000 sq. deg. sky was covered during the Twilight Survey between 2019 September to 2020 January. A sky coverage map of the 90 Twilight Survey observing sessions is presented in Fig.~1A. The apparent asymmetry between the morning and evening patches is due to differences in the accessibility of the sky in the Sun's direction during evening and morning twilight during the September to January months. The solar elongation of the sky covered by the Twilight Survey ranged between 35-60 degrees (Fig.~1B). Approximately half of the sky covered by the Twilight Survey is within the IVA maximum solar elongation range of $<$46 degrees (Fig.~1B). The near-Earth asteroid model \citep[][]{Granvik2018} predicts that $\sim$80$\%$ of IVAs have a maximum Solar elongation overlapping with the solar elongation range covered by the Twilight Survey (Fig.~1C).

\begin{figure}
\hspace{0 mm}
\centering
\includegraphics[scale = .25]{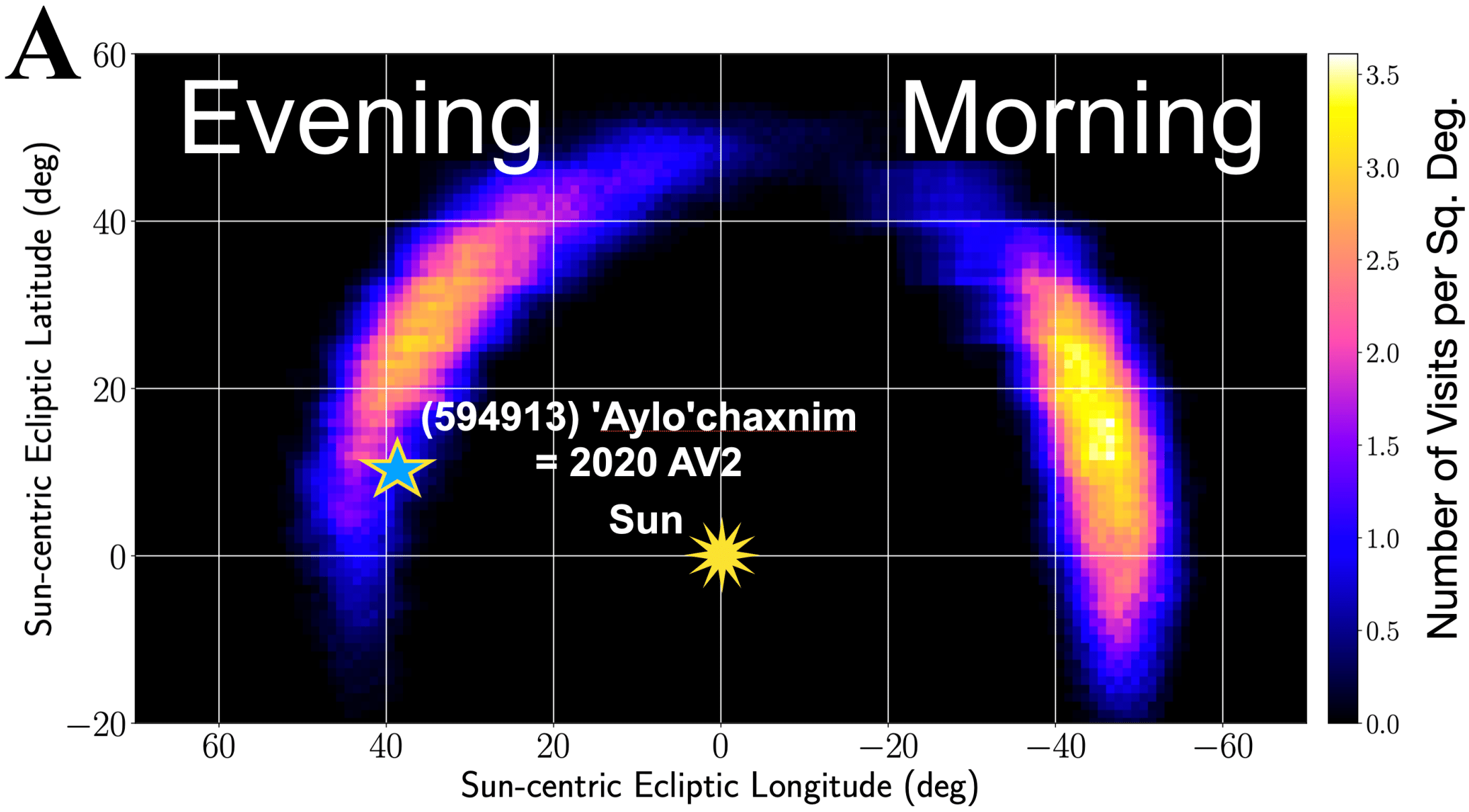}
\includegraphics[scale = 0.208]{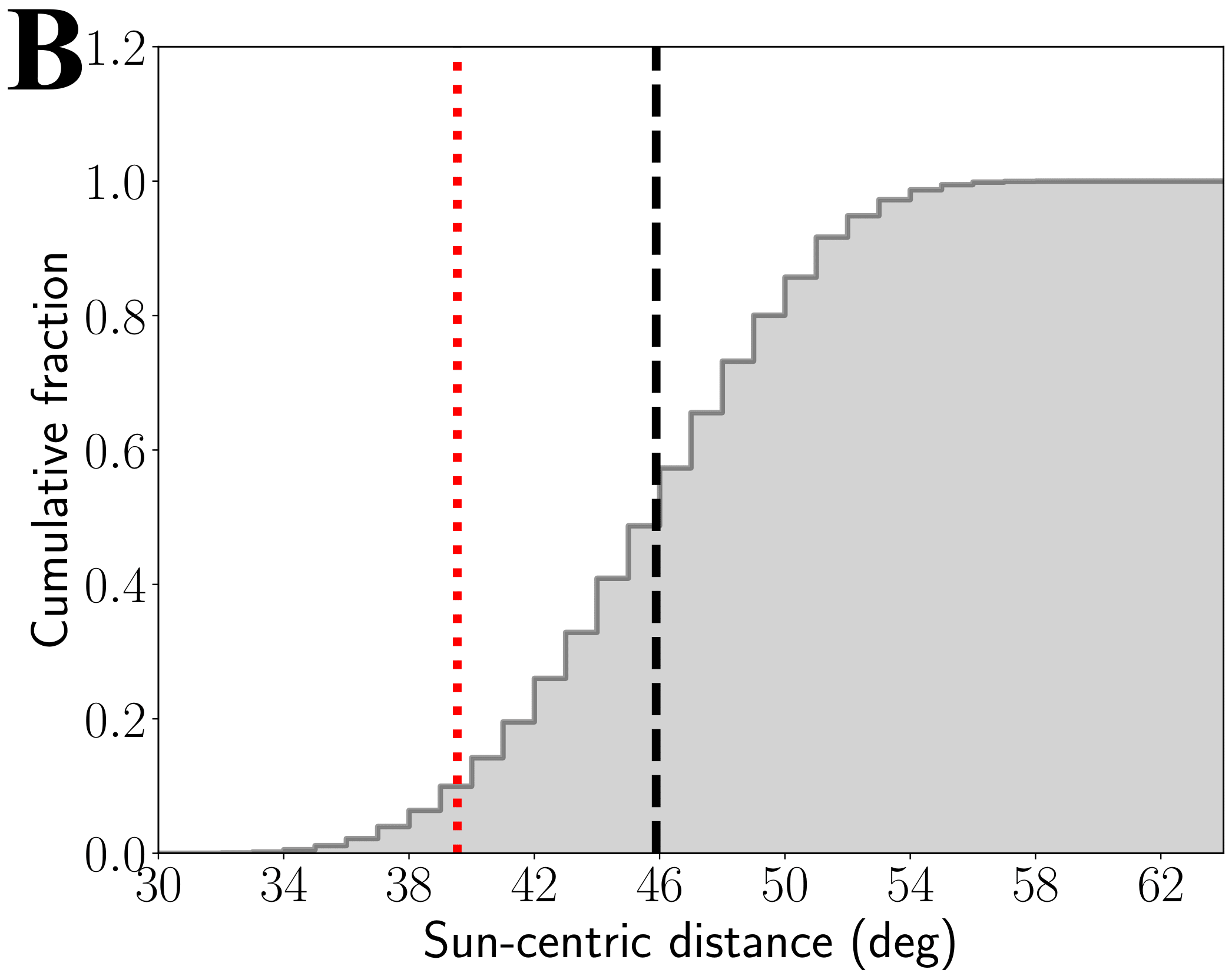}
\includegraphics[scale = 0.205]{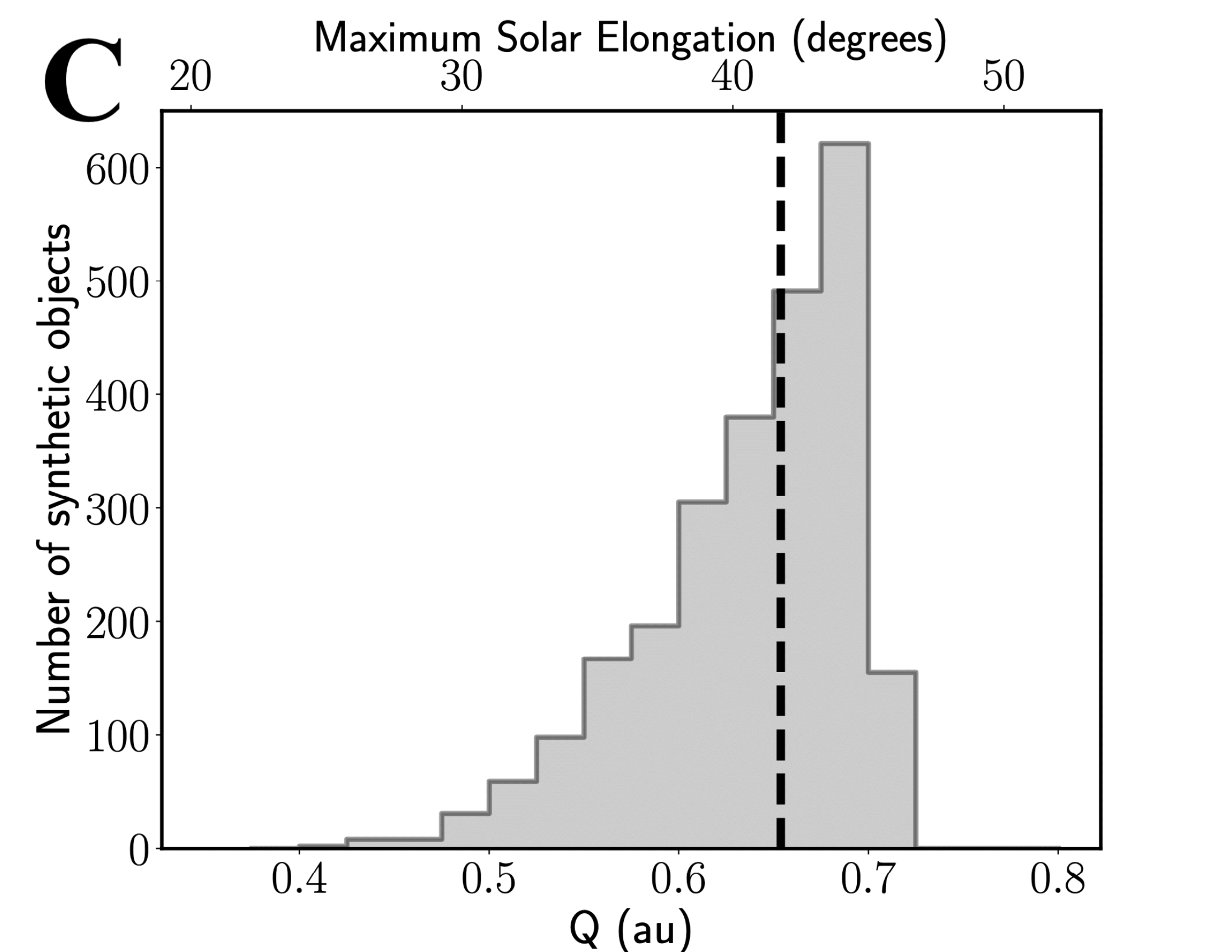}
\caption{ZTF Twilight Survey sky coverage. (A) Sun-centric sky distribution of Twilight Survey coverage between 2019 September 19 and 2020 January 30. The blue star indicates the first observed location of \an and the yellow star indicates the Sun. The colour scale is the number of ZTF visits per square degree. (B) Cumulative distribution of the Sun-centric distance of the Twilight survey footprints in the same time period. The vertical black dashed line shows the maximum possible Sun-centric distance of inner-Venus objects and the vertical red dotted line shows the Sun-centric distance of \an when it was first observed on 2022 January 4. (C) Maximum solar distance, $Q$, distribution using synthetic inner-Venus objects generated from the NEA model. The vertical dashed line shows the aphelion of \anns. The upper x-axis shows the maximum solar elongation of the synthetic inner-Venus object population.}
\end{figure}

\section{Results}
\subsection{Initial detection}
On 2020 January 4, \av was detected by ZTF in the evening twilight sky $\sim$40 degrees from the Sun (Fig.~2A,B). Follow-up data were obtained with the Kitt Peak Electron Multiplying Charge-Coupled Device Demonstrator (KPED) mounted on the Kitt Peak 84-inch telescope \citep[][]{Coughlin2019} on 2020 January 9 and were reported to the Minor Planet Center (MPC). The astrometry from the follow-up observations combined with the initial ZTF observations refined the aphelion, $Q$, of \av as having a value of $\sim$0.65 astronomical units (au), well within the 0.72~au perihelion, $q$, of Venus, as seen in the top and bottom panels of Fig.~3, \citep[][]{Bolin2020MPEC}.

\begin{figure}\centering
\includegraphics[width=.4\linewidth]{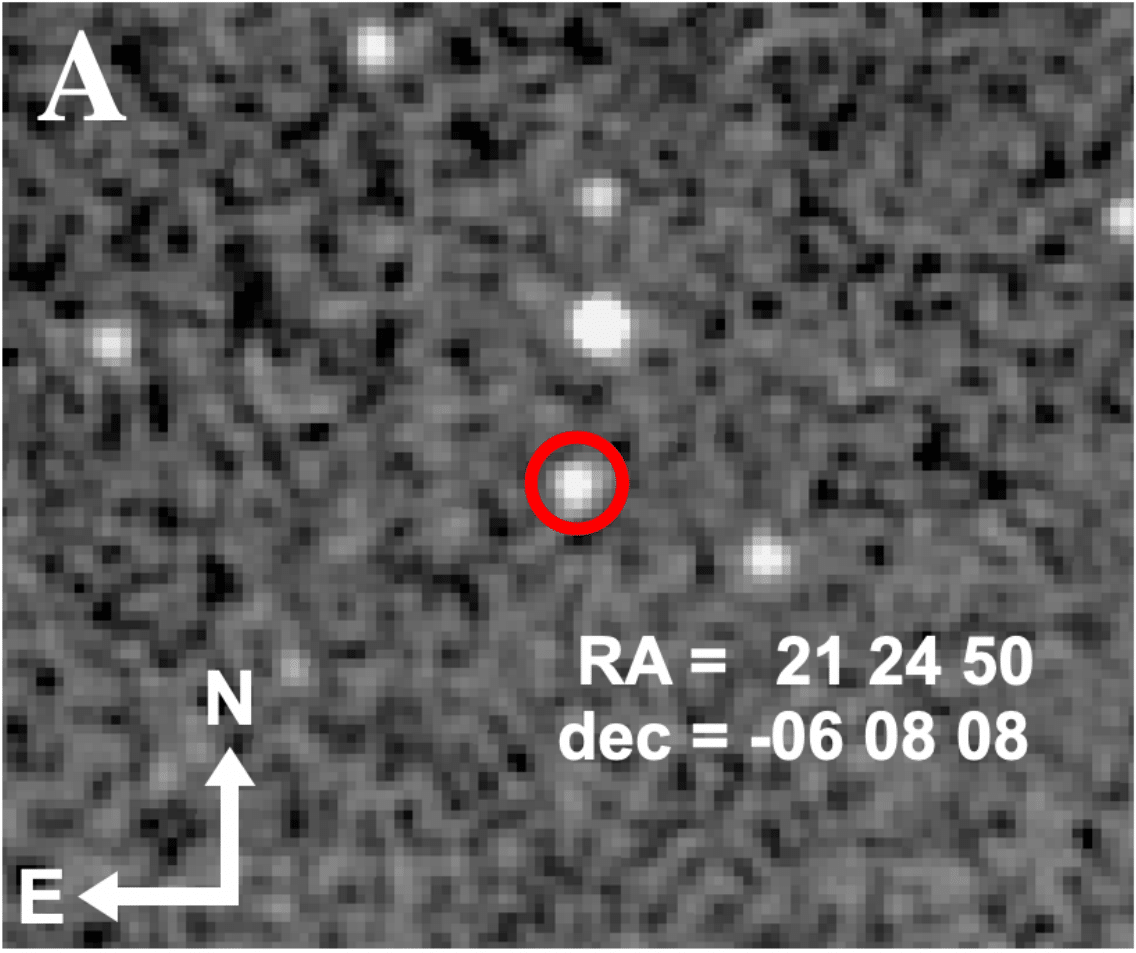} 
\includegraphics[width=.4\linewidth]{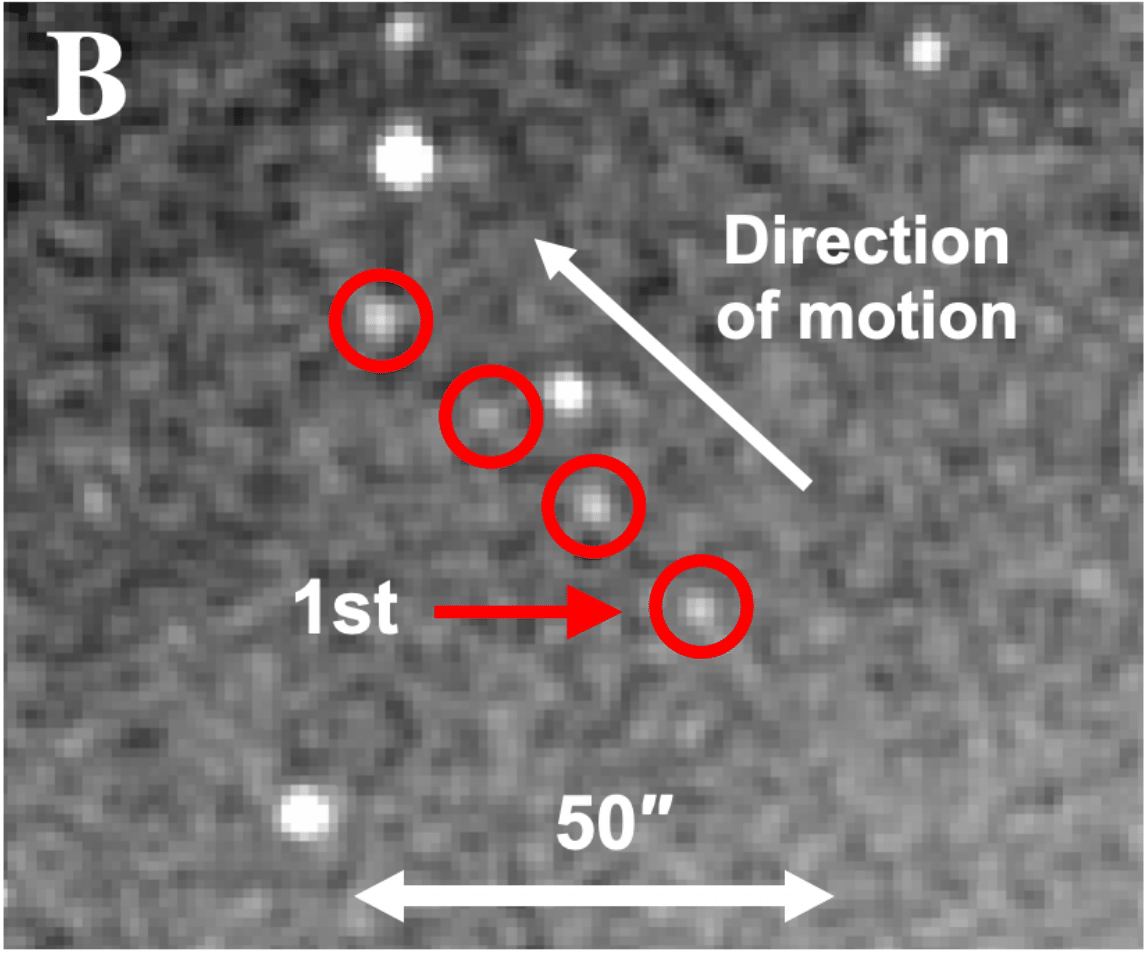}
\includegraphics[width=.4\linewidth]{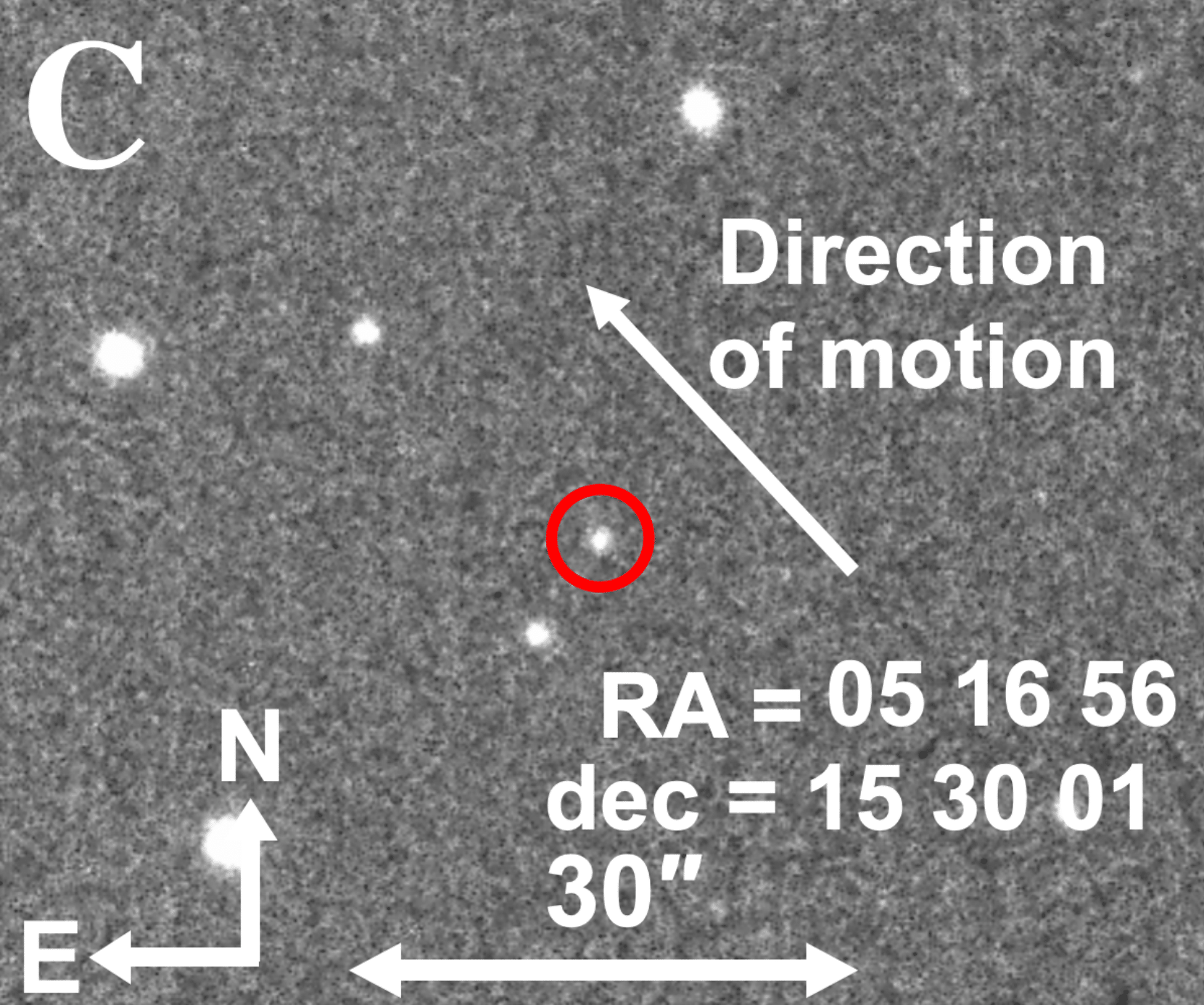} 
\caption{ZTF images of \anns. (A) One of the ZTF discovery 30~s r-band images of \an taken on 2020 January 4. \an is indicated by the red circle. (B) Composite image of the first four ZTF 30~s r-band exposures over a 22 minute time interval. The images were aligned on the background stars before being coadded. \an is indicated by the red circles, with the first observation labelled with a red arrow. The apparent faintness of the third detection is due to variations in sky transparency over the 22 minute sequence. The asteroid was moving $\sim$130 arcseconds per hour in the northeast direction resulting in a $\sim$10 arcseconds spacing between the detections of \anns. The spatial scale in (B) is the same as in (A). (C) Recovery image of \an made using the 6.5~m Magellan Baade telescope and Four Star infrared camera on 2021 July 18. The asteroid was moving in the northeast direction at 100 arcseconds per hour. The cardinal directions are indicated.}
\end{figure}

Follow-up data from other observatories were reported during 2020 January 4-23 resulting in a more precise orbit fit with $Q$ = 0.653817$\pm$0.000825~au. The orbit of \av was further refined when recovery observations were taken by follow-up observers during its next window of observability from the northern hemisphere in 2020 November 24-26 and reported to the MPC\footnote{\url{https://minorplanetcenter.net/db_search/show_object?object_id=K20A02V}}. These additional recovery observations extended the observing span to 327 days improving the precision of \avns's orbital elements to one part in a million \citep[Table~A1,][]{Ip2021supp}. A third set of observations confirming the orbit were obtained in 2021 July 17-19 at the Southern Astrophysical Research Telescope and Magellan Telescope \citep[Fig.~2C,][]{Ip2021supp}. The more precise orbit fit enabled by these three epochs of observations resulted in \av receiving the number designation (594913) by the MPC on 2021 September 20 \citep[][]{AV2MPC}. On 2021 November 8, the International Astronomical Union Working Group on Small Body Nomenclature officially named the asteroid \anns, meaning ``Venus Girl'' in the Luise\~{n}o language \citep[][]{Ticha2021}. We suggest that the class of interior to Venus asteroids be referred to as \an following the example of (594913) \an as the first known example of this class of asteroids.

\begin{figure}
\centering
\includegraphics[scale = .29]{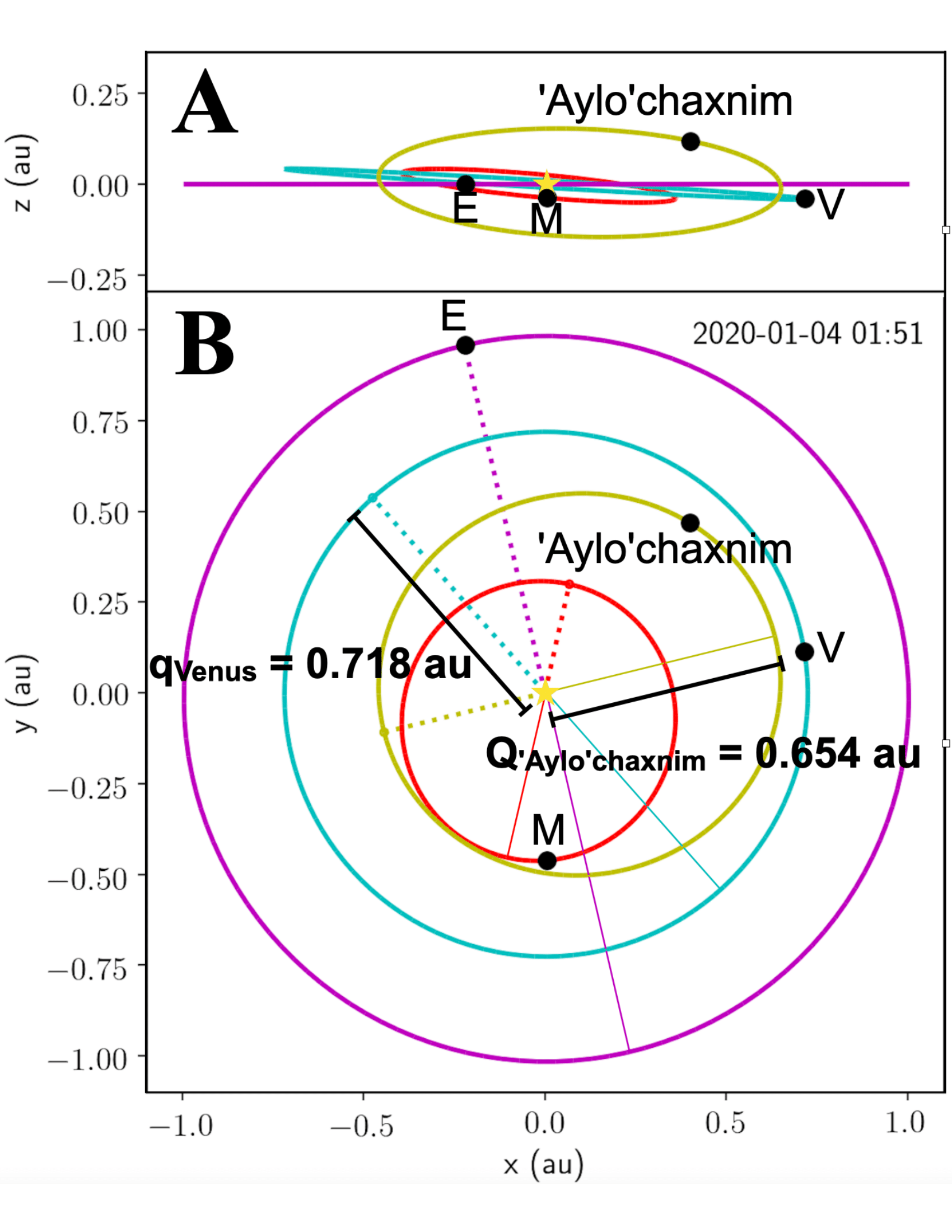}
\caption{Orbital configuration of \anns. (A) A side view of the plane of the Solar system. The orbits and locations of Earth (purple), Venus (blue), Mercury (red) and \an (yellow) on 2020 January 4 are shown looking from the side. (B) The same as panel (A) but looking from above the orbital plane of the inner Solar System. The perihelion and aphelion directions of \an and the planets are plotted with dotted and solid lines respectively.The perihelion distance of Venus, $q_{\mathrm{Venus}}$, and the aphelion distance of \anns, $Q_{\mathrm{\textquotesingle Aylo\textquotesingle chaxnim}}$ are indicated with labeled barred lines. The heliocentric cartesian coordinates $x$, $y$ and $z$ are indicated with the position of the Sun as the origin.}
\end{figure}

\subsection{Orbital dynamics}
Previous simulations of the orbital evolution of \an \citep[][]{Greenstreet2020,FuenteMarcos2020} indicate its capture into orbital period resonances with Venus, such as the 3:2 mean motion resonance located at 0.552~au. We performed further orbital stability simulations using the orbital solution of \an from 2020 November, finding that the nominal orbit of \an enters the 3:2 mean motion with Venus $\sim$0.06~Myr from now (Fig.~4A). The amplitude of the variations in the semi-major axis, $a$, of \an is initially large but shrinks after close encounters with Mercury $\sim$0.01~Myr later.  The minimum approach distance between \an and Venus increases when \an enters into this resonance, thereby protecting the asteroid from close encounters with the planet \citep[similar to the 3:2 mean motion resonances between Neptune and Pluto][]{Nesvorny2000}. The integration of the majority of our orbital clones indicates that the asteroid \an will remain in resonance with Venus for $\sim$ 0.01~Myr, and subsequently leave and re-enter the 3:2 mean motion resonance for the next $\sim$0.1~Myr.

\begin{figure}
\centering 
\hspace{-0.8 mm}
\includegraphics[scale=0.308]{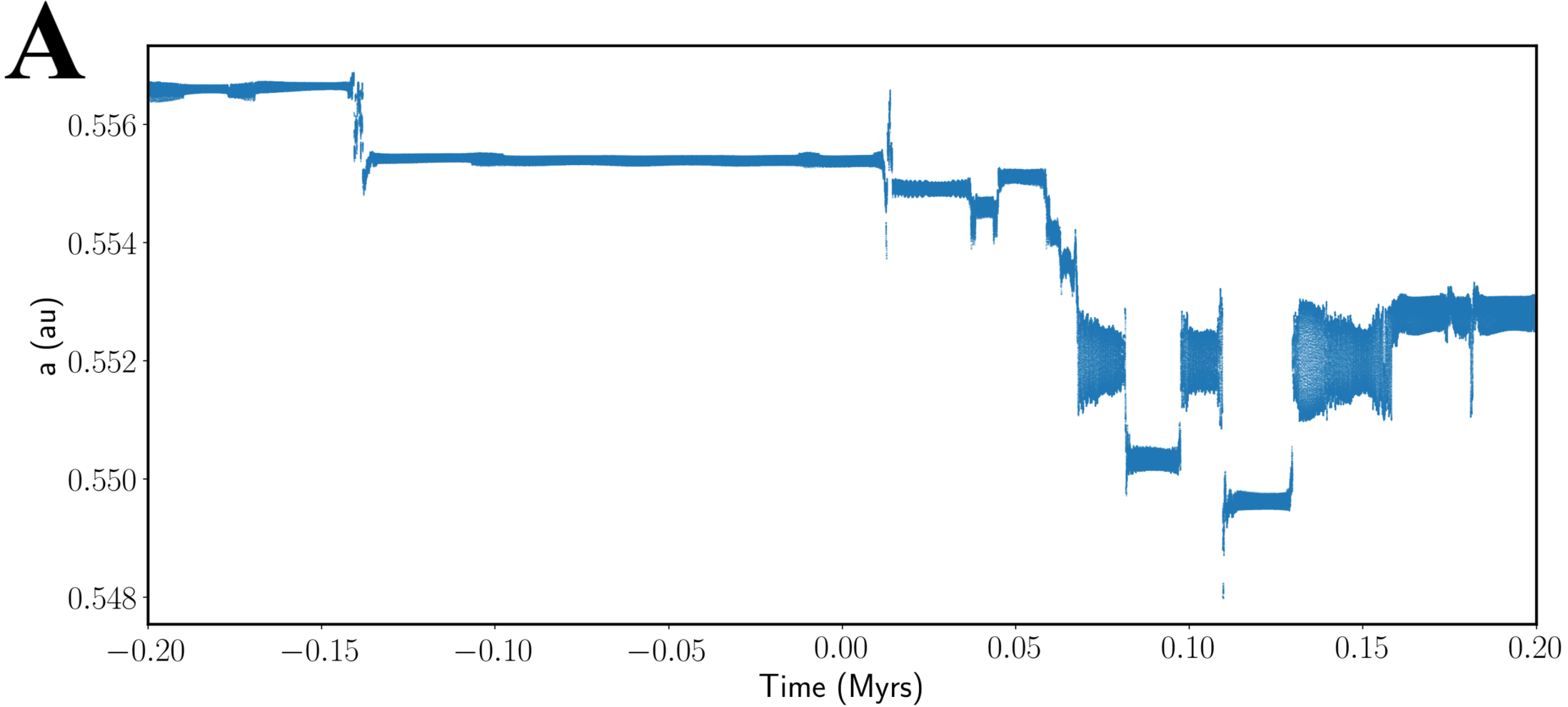}
\includegraphics[scale=0.308]{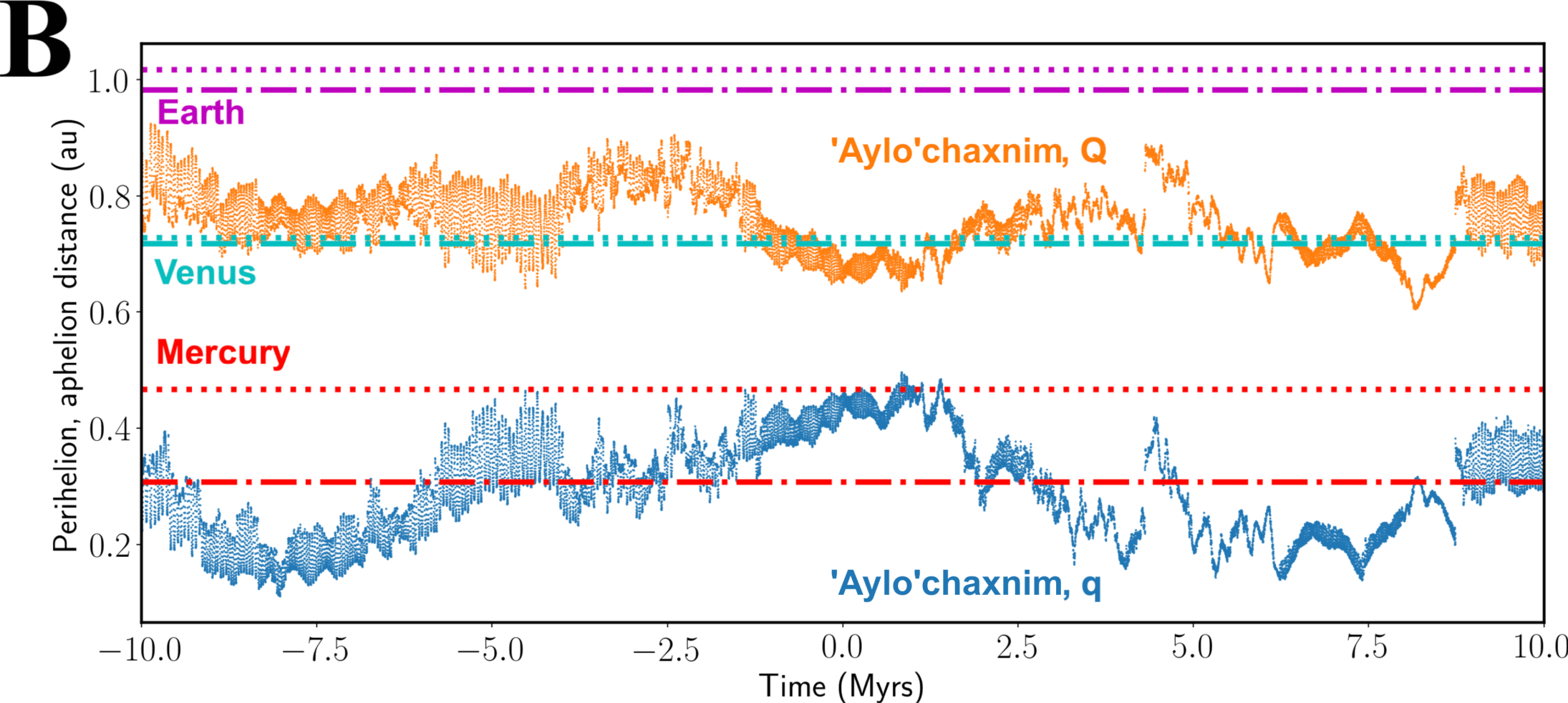}
\includegraphics[scale=0.304]{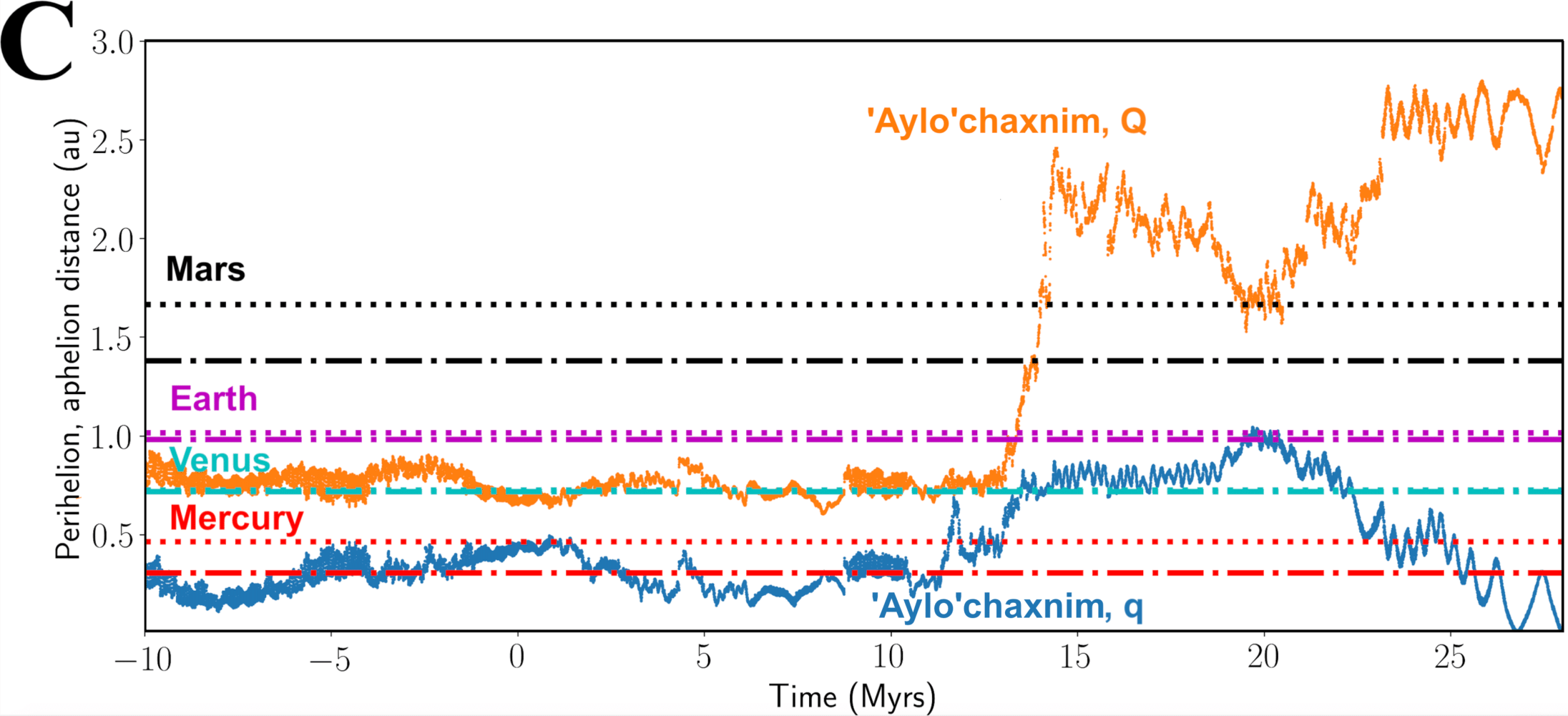}
\caption{Orbital evolution of a synthetic clone representing the nominal orbit of \anns. (A) Evolution of the semi-major axis of \an represented by its nominal orbit in Table~A1. The plateaus in the evolution of the semi-major axis separated by jumps are due to \an crossing different resonances with Venus. At around 0.06~Myrs, \an entered a 3:2 mean motion resonance with Venus located at 0.552~au that lasts for a 0.01~Myrs before jumping in out of the resonance for the next $\sim$0.1~Myrs. (B) The evolution of the aphelion (orange) and perihelion (blue) distances of the same clone of \an representing the nominal orbit integrated to $\pm$10~Myrs. The current aphelion (dashed line) and perihelion distances (dash-dot line) are plotted as horizontal lines for Mercury (red), Venus (cyan) and Earth (purple). (C) Similar to B, but for a selected long-lived clone integrating its orbital evolution to 28~Myrs. The aphelion and perihelion range of Mars is shown in black.}
\end{figure}

Our orbital evolution simulations also indicate that \an has only recently migrated entirely inside the orbit Venus, within the last $\sim$1~Myr, remaining inside the orbit of Venus for another $\sim$2~Myr (Fig.~4B). The proximity of the perihelion of \an to the orbit of Mercury draws comparison with the perihelia of many ecliptic comets being close to the orbit of Jupiter, indicative of the evolution of their orbits due to close planetary encounters \citep[][]{Duncan2004}. The orbit of \an will have aphelia within the orbit of Venus for the next 2~Myr. Previous simulations found a much shorter residence time of $<$1~Myr for \an to remain inside the orbit of Venus \citep[][]{Greenstreet2020,FuenteMarcos2020}. We ascribe the differences with our results as due to previous work adopting an earlier version of the orbital parameters for \anns.

While the current precision of \anns's orbit prevents us from predicting its orbital behavior on timescales exceeding $\sim$10~Myr, it is apparent from orbital integration of clones \citep[see Sec.~A1.3, ][]{Ip2021supp} of its orbit that it is a transitory inhabitant of the inner Venus region of the Solar System. The majority of orbital clones have close encounters with Mercury, Venus and the Earth within 10-20~Myr that scatter and evolve their orbits onto excited trajectories that have very close perihelion passages with the Sun (Fig.~4C). The median time between the start time of the integration of the \an clones and their collision with a planet or the Sun is $\sim$10~Myr, and $\sim$90$\%$ of the clones have collided with the Sun or a planet by the end of the 30~Myr integration. We integrated the $\sim$10$\%$ of clones that survived the first 30~Myr for a total of 50~Myr. On that time-scale, $\sim$13$\%$ of the \an clones collided with the Sun, having a perihelion distance $<$0.005~au, while 13$\%$, 52$\%$, 16$\%$ and 2$\%$ collided with Mercury, Venus, the Earth and Mars, respectively. The remaining 4$\%$ survived the extended 50~Myr integration or were ejected from the Solar System. Previous simulations of inner-Venus objects found a median collisional lifetime of $\sim$21~Myr \citep[][]{Greenstreet2012vat}, a factor of 2 larger than we find for \an clones. However, the proportion of \an clones colliding with the Sun, Mercury, Venus, Earth and Mars is similar to the published simulations of the general inner-Venus object population \citep[][]{Greenstreet2012vat}.

\subsection{Spectral type, dynamical pathway and size}
Spectroscopic observations of \an were made using the Keck telescope on 2020 January 23 indicating a reddish surface with colours in equivalent $g$ (wavelength $\sim$470~nm), $r$ (wavelength $\sim$620~nm) and $i$ (wavelength $\sim$750~nm) bandpasses of $g$-$r$ = 0.65$\pm$0.02 mag, $r$-$i$ = 0.23 $\pm$ 0.01~mag. In addition, the surface of \an has a $i$-$z$ colour of 0.11$\pm$0.02~mag where the equivalent $z$ bandpass corresponds to a central wavelength of $\sim$900~nm (Fig.~5). We interpret these as indicating a silicate S-type asteroid-like composition \citep[][]{Bus2002}, consistent with an origin from the inner Main Belt, where S-type asteroids are most abundant \citep[][]{Demeo2013}. NEA models predict that asteroids with similar orbital elements to \an \citep[Table~A1, Fig.~A1,][]{Ip2021supp} originate from the inner Main Belt \citep[][]{Granvik2018}.

One of the possible dynamical pathway for inner-Venus asteroids (IVAs) is to originate from the Main Belt through source regions located near various major planetary resonances \citep[][]{Granvik2017}. If we assume that \an originated from the Main Belt as an asteroid family fragment \citep[][]{Bolin2017} before crossing inside of the orbit of Venus, asteroids with orbits similar to \an most likely originate from the $\nu_6$ resonance, with a $\sim$77$\%$ probability, that forms the boundary of the inner Main Belt at 2.2~au \citep[][]{Morbidelli1994,Granvik2018}. The second most likely source of \an with a $\sim$18$\%$ probability are the Hungaria asteroid population located just interior to the Main Belt at 2.0~au \citep[][]{Milani2010a} and the third most likely at $\sim$4$\%$ being the 3:1 mean motion resonance with Jupiter located in the Main Belt at 2.5~au \citep[][]{Wisdom1983}.

The typical albedo value for S-type asteroids is $\sim$0.2 \citep[][]{Thomas2011,Demeo2013}. In addition, we compared the source region probability for \an with the medium-resolution version of the NEA albedo \citep[][]{Morbidelli2020}. The NEA albedo model predicts that $\sim$60$\%$ of kilometre-size inner-Venus objects should have albedos exceeding 0.2 \citep[see Fig.~A3, ][]{Ip2021supp} consistent the albedo based on \anns's taxonomic type. For an albedo of 0.2 and the absolute magnitude of \anns, $H$ = 16.2$\pm$0.8~mag taken from from the JPL Small Body Database\footnote{\url{https://ssd.jpl.nasa.gov/tools/sbdb\_lookup.html\#/?sstr=594913}}, we estimate that \anns's diameter is $\sim$1.7$\pm$0.6~km.

 The number of IVAs in the NEA model brighter than \anns's nominal value of $H$ = 16.2 is 0.25. Using the range of $H$ described by its 1-$\sigma$ uncertainty of 0.8, the number of objects brighter than the 1-$\sigma$ lower value of $H$ = 15.4 is 0.05, and the number of objects brighter than the 1-$\sigma$ upper value of $H$ = 17.0 is 0.7. Thus, the number of objects brighter than $H$ $<$ 16.2$\pm$0.8 is 0.25$\pm$$^{0.45}_{0.20}$ with the main source of uncertainty in the number of objects being due to the uncertainty on the $H$ magnitude of \anns. Thus, using the 1-$\sigma$ upper uncertainty value on $H$, and without taking into account small number statistics or observational selection effects, there is a $\sim$1-2-$\sigma$ difference between the discovery of \an and the NEA model.
 
\begin{figure}
\hspace{-0 mm}
\includegraphics[scale=0.21]{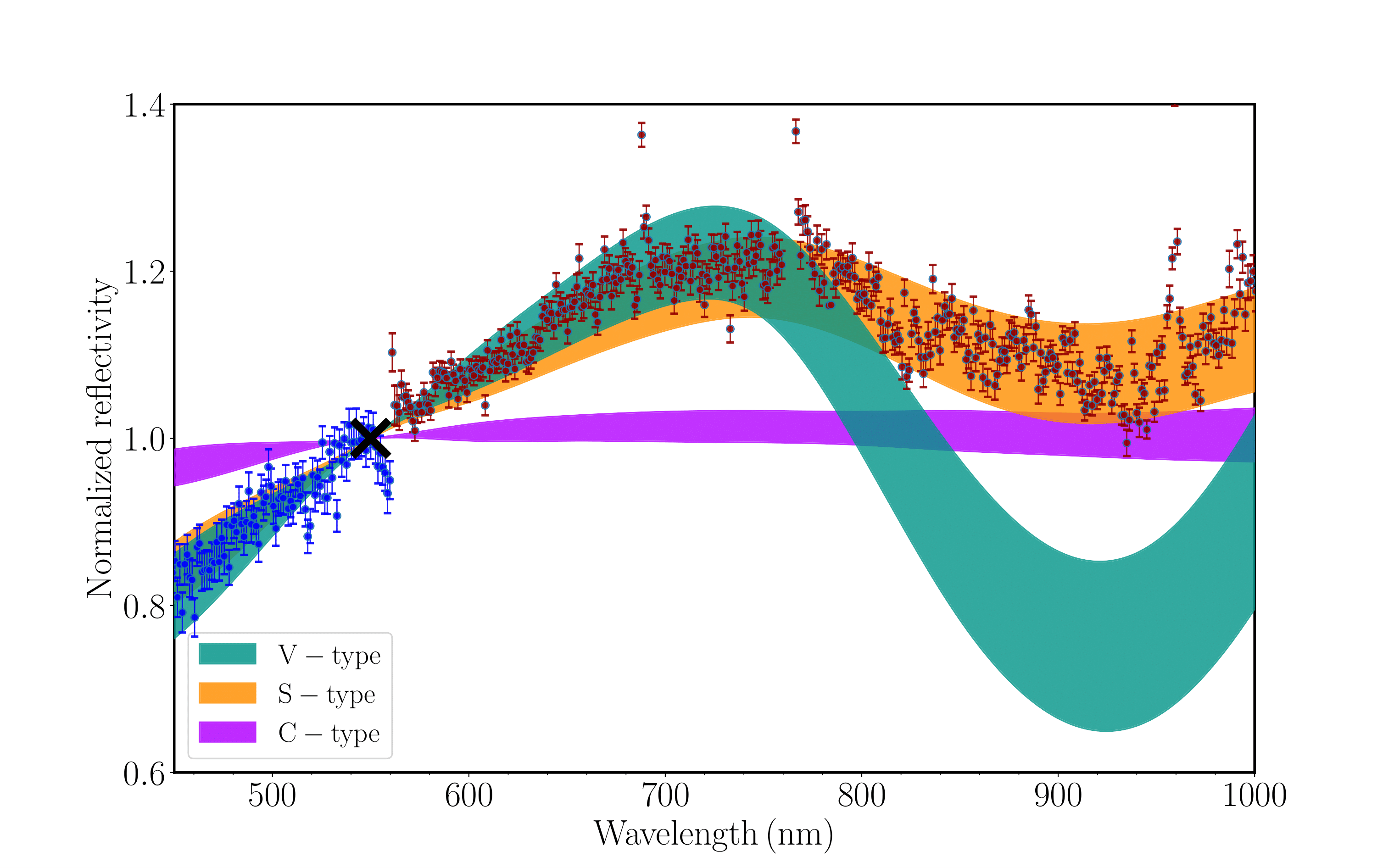}
\caption{Visible wavelength reflectance spectrum of \anns. Taken with the LRIS instrument on Keck I on 2020 January 23, the spectrum of \an is plotted as blue dots with 1-$\sigma$ uncertainty error bars. The spectrum has been normalised to unity at 550~nm indicated by the black cross. The spectrum was obtained by combining two spectra from the blue camera (blue data points) and the red camera (red data points). The data have been rebinned by a factor of 10 using an error-weighted mean. The dip at $\sim$560~nm, spikes at $\sim$770~nm and $\sim$960~nm are artefacts caused by the dichroic and imperfect removal of telluric H$_2$O absorption features. The spectral range of S, V and C-type asteroids are over-plotted.}
\end{figure}

\section{Discussion and conclusion}
\an seems like an ordinary NEA with a red colour and orbital evolution affected by planetary encounters. However, while asteroid population models predict that there are $\sim$1,000 km-scale NEAs, IVAs are scarce, representing less than 0.3$\%$ of the NEA population \citep[][]{Granvik2018, Morbidelli2020}. The detection of \an is surprising given its large size and the relative rarity of IVAs according to the NEA model. However, the twilight sky within 50 degrees of the Sun is relatively unexplored and and the comparison between observations and asteroid population models requires future exploration of this phase space. Observations of the near-Sun sky by current surveys such as ZTF and the Dark Energy Camera \citep[][]{Sheppard2021DPS} along with future surveys such as the Rubin Observatory Legacy Survey of Space and Time \citep[][]{Bianco2022} will provide coverage of the near-Sun sky and the IVA population.

\section*{Acknowledgements}
The authors appreciate the help of Frank Masci with ZTF asteroid identification, Alessandro Morbidelli with sythetic populations, Jacqueline Ser\'{o}n, Carlos Corco and Alfredo Zenteno with SOAR observations, Peter Senchyna, Alan Dressler, Carla Fuentes, Carlos Contreras and Andy Monson with Magellan observations, R. Quimby, M. W. Coughlin and K. B. Burdge, M.J. Graham with follow up. C.F.~acknowledges support from the Heising-Simons Foundation (grant $\#$2018-0907). Part of this work was carried out at the Jet Propulsion Laboratory, California Institute of Technology, under contract with NASA 80NM0018D0004.
\section*{Data Availability}
The data underlying this article will be shared on reasonable request to the corresponding author. The Twilight Survey data from 2019 September and 2020 January are available in ZTF Public Data Release 7.

\section*{Supplemental Material}
The supplemental material for this manuscript is available online.



\bibliographystyle{mnras}
\bibliography{../scibib} 




\renewcommand{\thefigure}{A\arabic{figure}}
\setcounter{figure}{0}
\renewcommand{\thetable}{A\arabic{table}}
\renewcommand{\theequation}{A\arabic{equation}}
\renewcommand{\thesection}{A\arabic{section}}
\setcounter{section}{0}
\setcounter{page}{1}
\renewcommand\thepage{A\arabic{page}}
\section*{Supplemental Material}
\appendix
\renewcommand{\thefigure}{A\arabic{figure}}
\setcounter{figure}{0}
\renewcommand{\thetable}{A\arabic{table}}
\renewcommand{\theequation}{A\arabic{equation}}
\renewcommand{\thesection}{A}
\setcounter{section}{0}
\subsection{Discovery, follow-up and orbital determination}
The initial discovery observations of \an on 2020 January 4 were made in four sidereally-tracked 30~s $r$-band exposures with the 48-inch Samuel Oschin Telescope. The observations were made during evening astronomical twilight while the telescope was pointing at 25.5 degrees elevation and the center of the telescope's field of view was pointing through 2.3 airmasses. The brightness of \an during its discovery observations was $r$$\sim$18.1 magnitude and was moving approximately $\sim$130~arcseconds per hour resulting in negligible trailing losses of the individual detections of \an in the images (Fig.~2, A-B). The observations were taken during average seeing conditions for the Palomar site with nearby stellar objects of similar brightness to \an having a full width at half maximum (FWHM) of $\sim$2.1~arcseconds. The effect of airmass on the astrometric uncertainties of observations at the Palomar Observatory site are small, increasing to $<$0.08-0.10 arcseconds in declination at the 2.0-3.0 airmass range of the observations of \an \citep[][]{Masci2019}. Twilight observing sessions were conducted each night, alternating between evening and morning twilight on sequential nights. The typical limiting magnitude in a single Twilight Survey exposure is V$\sim$20 as seen in Fig.~A1. The orbital elements based on observations of \an taken between 2020 January 4 and 2020 November 26 is listed in Table~A1 and Fig.~A2.

\begin{figure}
\centering 
\includegraphics[width=0.8\linewidth]{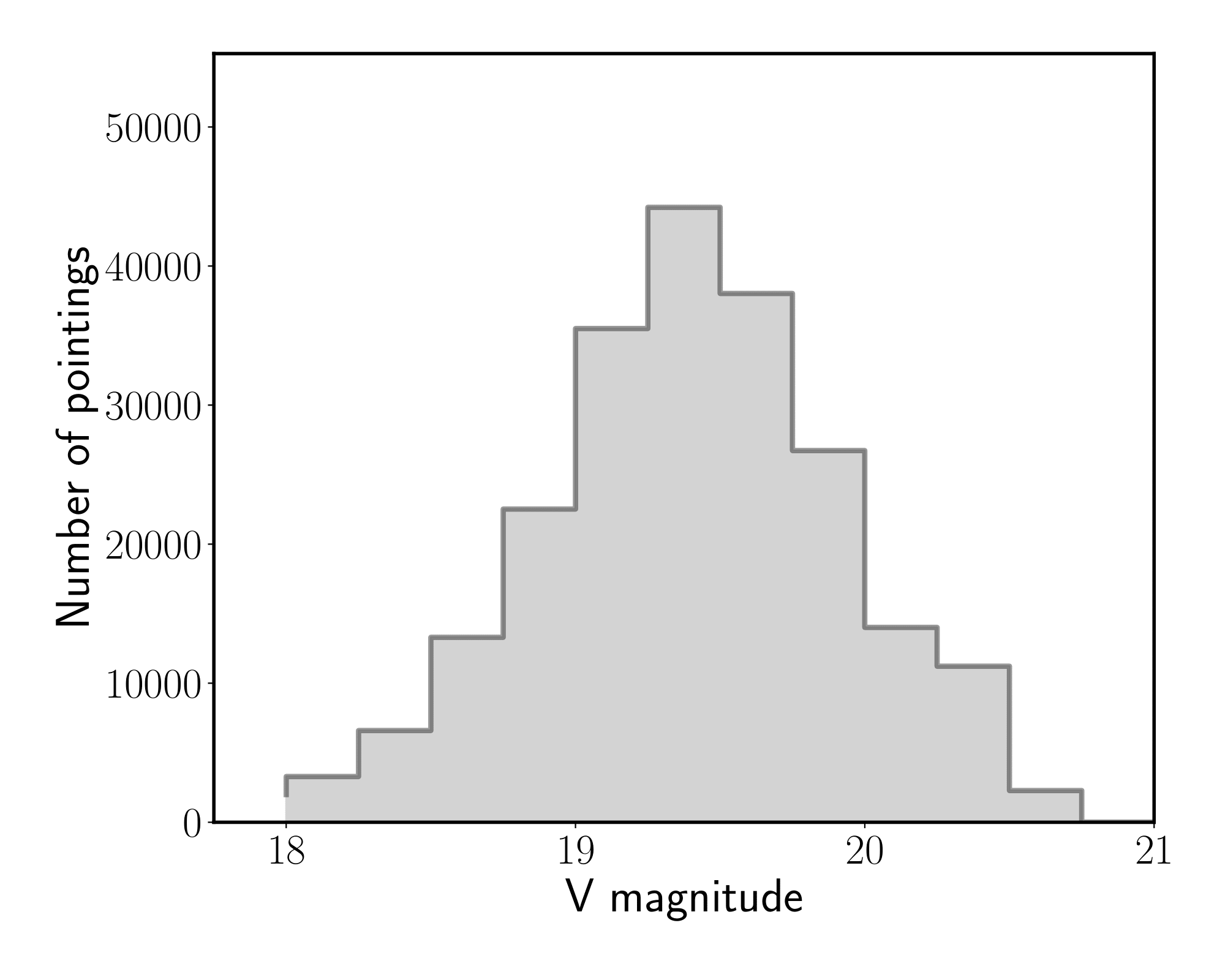}
\caption{The exposure-based 5-$\sigma$ limiting magnitude of the Twilight survey pointings between 2019 September 20 and 2020 January 30.}
\end{figure}

\begin{table}
\centering
\caption{Orbital elements of (594913) \an based on observations reported to the MPC taken between 2020 January 4 to 2020 November 26. The orbital elements are shown for the Julian date (JD) 2,459,179.5. The 1-$\sigma$ uncertainties are given in parentheses. The value and 1-$\sigma$ uncertainties for $H$ were taken from the JPL Small Body Database on 2022 July 14.}
\label{t:hstobs}
\begin{tabular}{ll}
\hline
Element&
\\ \hline
Epoch (JD) & 2,459,179.5\\
Time of perihelion, $T_p$ (JD) & 2,458,907.045$\pm$(0.019)\\
Semi-major axis, $a$ (au) & 0.555443170$\pm$(3.65x10$^{-7}$)\\
Eccentricity, $e$ & 0.17696610$\pm$(2.33x10$^{-6}$)\\
Perihelion, $q$ (au) & 0.45714852$\pm$(1.56x10$^{-6}$)\\
Aphelion, $Q$ (au) & 0.653737830$\pm$(9.53x10$^{-7}$)\\
Inclination, $i$ ($^{\circ}$) & 15.86824$\pm$(0.00014)\\
Ascending node, $\Omega$ ($^{\circ}$) & 6.70827$\pm$(0.00035)\\
Argument of perihelion, $\omega$ ($^{\circ}$) & 187.32773$\pm$(0.00053)\\
Mean Anomaly, $M$ ($^{\circ}$) & 288.6292$\pm$(0.0011)\\
Absolute Magnitude, $H$ & 16.2$\pm$(0.8)\\
\hline
\end{tabular}
\end{table}

\begin{figure}
\includegraphics[scale = .42]{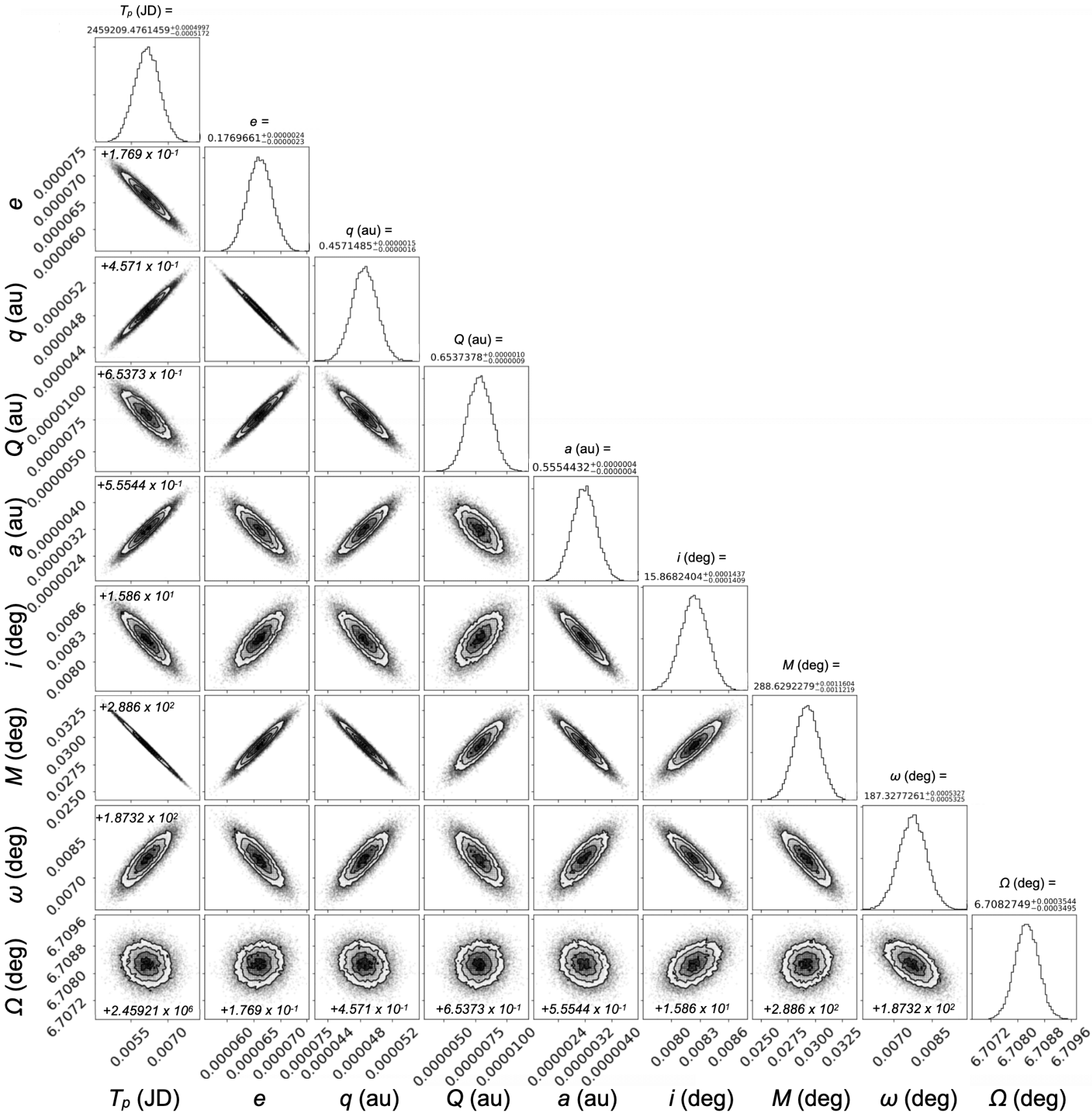}
\caption{Distribution of \an orbital element samples. Corner plot of 30,000 samples from the multivariate distribution of orbital elements of \anns. The samples were derived from the covariance by fitting the observations between 2020 January 4 to 2020 November 26. Two dimensional histograms are shown for each pair of combinations of $T_p$, $e$, $q$, $Q$, $a$, $i$, $M$, $\omega$ and $\Omega$ for the orbital samples. Individual samples are plotted as tiny grey dots. The contours represent the 0.5,1, 1.5, 2-$\sigma$ levels for a two dimensional histogram.  One dimensional histograms are shown for the $T_p$, $e$, $q$, $Q$, $a$, $i$, $M$, $\omega$ and $\Omega$ values of the orbital samples. The central value and the 1-$\sigma$ uncertainty for each parameter value is given at the top of each column in the corner plot.}
\end{figure}

\subsubsection{Discovery, follow up and characterization observation details}
\textit{Zwicky Transient Facility, ZTF:} The ZTF camera consists of 16 separate 6144~pixel $\times$ 6160~pixel arrays on a single CCD camera mounted on the 48-inch Samuel Oschin Telescope at Palomar Observatory and is robotically operated. The plate scale of the camera is 1.01 arcseconds pixel$^{-1}$ and has a 7.4~degree $\times$ 7.4~degree field of view \citep[][]{Bellm2019,Graham2019,Dekany2020}. The data processing pipeline produced images differenced from reference frames and removes or masks most detector artefacts. Transients are extracted from the images and several algorithms are used to identify slower moving objects that appear as round point spread function detections in the images \citep[][]{Masci2019}. 
\\
\textit{Kitt Peak Electron Multiplying CCD Demonstrator, KPED:} The KPED instrument was mounted on the Kitt Peak 84-inch telescope and consists of a 1024 $\times$ 1024 pixel Electron Multiplying CCD camera and was robotically operated \citep[][]{Coughlin2019}. The camera has a spatial scale of 0.26 arcseconds pixel$^{-1}$ and a 4.4~arcminute $\times$ 4.4~arcminute field of view. The camera is capable of reading out at a rate of 1 Hz and of individual exposures times up to 10 s. Our observations used 10\,s exposures in $r$-band and were sidereally tracked due to the short exposure time. Seeing conditions were $\sim$1 arcsec as measured for background stars in the follow-up images.
\\
\textit{Keck I Telescope:} The Low Resolution Imaging Spectrometer (LRIS) \citep[][]{Oke1995} on the Keck I telescope was used to observe \an on 2020 January 23 in spectroscopy mode (Program ID C272, PI M. Graham). The blue camera consists of two 2k $\times$ 4k Marconi CCD arrays and the red camera consists of two science grade Lawrence Berkeley National Laboratory 2k $\times$ 4k CCD arrays. Both cameras have a spatial resolution of 0.135 arcsec pixel$^{-1}$. The 1.0-arcsecond wide slit was used with the 560 nm dichroic with $\sim$50$\%$ transmission efficiency in combination with the 600/4000 grating for the blue camera and the 400/8500 grating for the red camera providing a spectral resolution of 0.4~nm and 0.7~nm, respectively \citep[][]{Oke1995,McCarthy1998}. A total exposure time of 600 s over two integrations were taken in seeing conditions of $\sim$0.6 arcseconds measured at zenith, however, the observations were taken at the large airmass of $\sim$3.4 resulting in degrading the seeing to $\sim$1.2 arcsec. Wavelength calibration used HgCdZn lamps for the blue camera and the ArNeXe lamps for the red camera. Flux calibration used the G191-B2B and Feige 34 standard stars for the blue and red camera respectively and a solar analog star 2MASS 22462446+0029244 was used for slope correction. The data of \an and the standard stars were obtained from the Keck Observatory Archive (KOA).
\\
\textit{Southern Astrophysical Research Telescope:} Observations of \an were conducted with the 4.1~m Southern Astrophysical Research Telescope (SOAR) on  2021 July 17 Goodman High Throughput Spectrograph in imaging mode \citep[][]{Clemens2004}. Data were taken in the SDSS $r$ filter, and the telescope was tracked non-sidereally at the asteroid's sky motion rate using 30~s exposures. The seeing during the observations was $\sim$1 arcsecond and the airmass of the observations was $\sim$3.3.
\\
\textit{Magellan Telescope:} The 6.5~m Magellan Baade telescope at Las Campanas Observatory was used to observe \an on 2021 July 18-19  using the Four Star infrared camera \citep[][]{Persson2013}. Non-sidereal tracking rates at the asteroid's sky motion rate were used and data were taken in the broadband $J$ filter in 4-9~s exposures using a four point dither pattern. The seeing during the observations was $\sim$1 arcsecond and the airmass of the observations was $\sim$2.4.
\\
\subsubsection{Visual imaging/spectroscopy reduction and astrometry}
The \textit{Gaia} data release 2 catalog \citep[][]{Gaia2016,Gaia2018} was used with the ZTF data reduction pipeline \citep[][]{Masci2019} to produce an astrometric solution on ZTF data with the \textsc{astrometrica} software\footnote{\url{http://www.astrometrica.at/}}. Photometric calibration was performed using the Pan-STARRS1 catalog database \citep[][]{Tonry2012}. The LRIS spectroscopic data were reduced using flat field, dark current and arc lamp exposures with the \textsc{lpipe} spectroscopy reduction software \citep[][]{Perley2019}. 
\\ 
\subsubsection{Orbit determination and dynamical integration}
We used the \textsc{rebound} N-body orbit integration package \citep[][]{Rein2012} using the \textsc{IAS15} adaptive time step integrator \citep[][]{Rein2015} to determine the orbital history of \anns. We generated 222 clone orbits of \an using the multi-variate distribution of the orbital parameters presented in Table~A1 and Fig.~A2, determined with observations taken between 2020 January 4 and 2020 November 26 with the \textsc{find$\_$orb} orbital integration package \footnote{\url{https://www.projectpluto.com/find_orb.htm}}. We integrated the \an orbital clones forwards and backward 30 Myrs using barycentric coordinates with a nominal timestep of 14~h. As an adaptive time-step integrator, \textsc{IAS15} decreases the time step during close encounters. We find that \an has experienced numerous, $\sim$0.01~au close encounters with Mercury and Venus that are as frequent as every few hundred to thousands of years due to it evolving on to planet-crossing orbits as seen for one example orbit clone of \an in Fig.~4C. We ignore the Yarkovsky effect in our orbital integrations because the effect of perturbations during planetary encounters on the orbital evolution of planet-crossing asteroids in the terrestrial planet zone dominates the Yarkovsky effect on the time scales of our orbital integrations \citep[][]{Bottke2002, Bottke2006, Granvik2018}.
\\
\subsection{Comparison with the near-Earth asteroid population and size estimate}
One of the main sources of uncertainty on the $\sim$1.7$\pm$0.6 km diameter estimate of \an is the 1-$\sigma$ uncertainty on $H$ of $\sim$0.8 magnitudes from JPL HORIZONS. We note that this uncertainty on the $H$ of \an is greater than the $\sim$0.3 magnitudes scatter on $H$ of asteroids from the Minor Planet Center catalog \citep[][]{Veres2015}. The large uncertainty on $H$ may be related to the unknown phase function \citep[][]{Jedicke2016}. We obtain a slightly lower diameter for \an of $\sim$1.4-1.6~km using albedos (in the range 0.23-0.3) measured for S-types in the general near-Earth Object (NEO)/near-Earth asteroid (NEA) population \citep[][]{Thomas2011}. The NEA albedo model \citep[][]{Morbidelli2020} predicts that $\sim60\%$ of km-scale inner-Venus objects should have albedos exceeding 0.2 as seen in Fig.~A3. Combining the spread in albedos of S-type asteroids \citep[][]{Thomas2011} with our estimate of the albedo of \an and the $\sim$0.8 magnitudes uncertainty on $H$, we estimate a diameter range for \an of $\sim$1.7$\pm$0.6~km.

The calculation of the absolute magnitude of \an was made assuming a phase parameter of 0.15. The assumed value of 0.15 is the average asteroid phase parameter determined by photometric observations of a large ensemble of asteroids regardless of spectral class \citep[][]{Oszkiewicz2012,Veres2015}.  Our spectrum and the spectrum reported by others of \an \citep[][]{Popescu2020} constrains the taxonomy of \an to be likely S-type (Fig.~5). Therefore, it might be appropriate to assume a higher phase parameter of $\sim$0.2 found for S-types \citep[][]{Pravec2012, Veres2015} which would increase the nominal absolute magnitude of \an by $\sim$0.1 magnitudes, smaller than the $\sim$0.8 absolute magnitude uncertainty. 

The scatter in phase parameters determined for asteroids within the S-types taxonomic class is $\sim$0.1-0.2 when measured over a large ensemble of asteroids \citep[][]{Warner2009, Veres2015}, and thus the nominal value for the phase parameter of $\sim$0.2 for S-type asteroids is compatible with our assumed phase parameter value of 0.15.

\begin{figure}
\centering 
\hspace{0 mm}
\includegraphics[scale = 0.32]{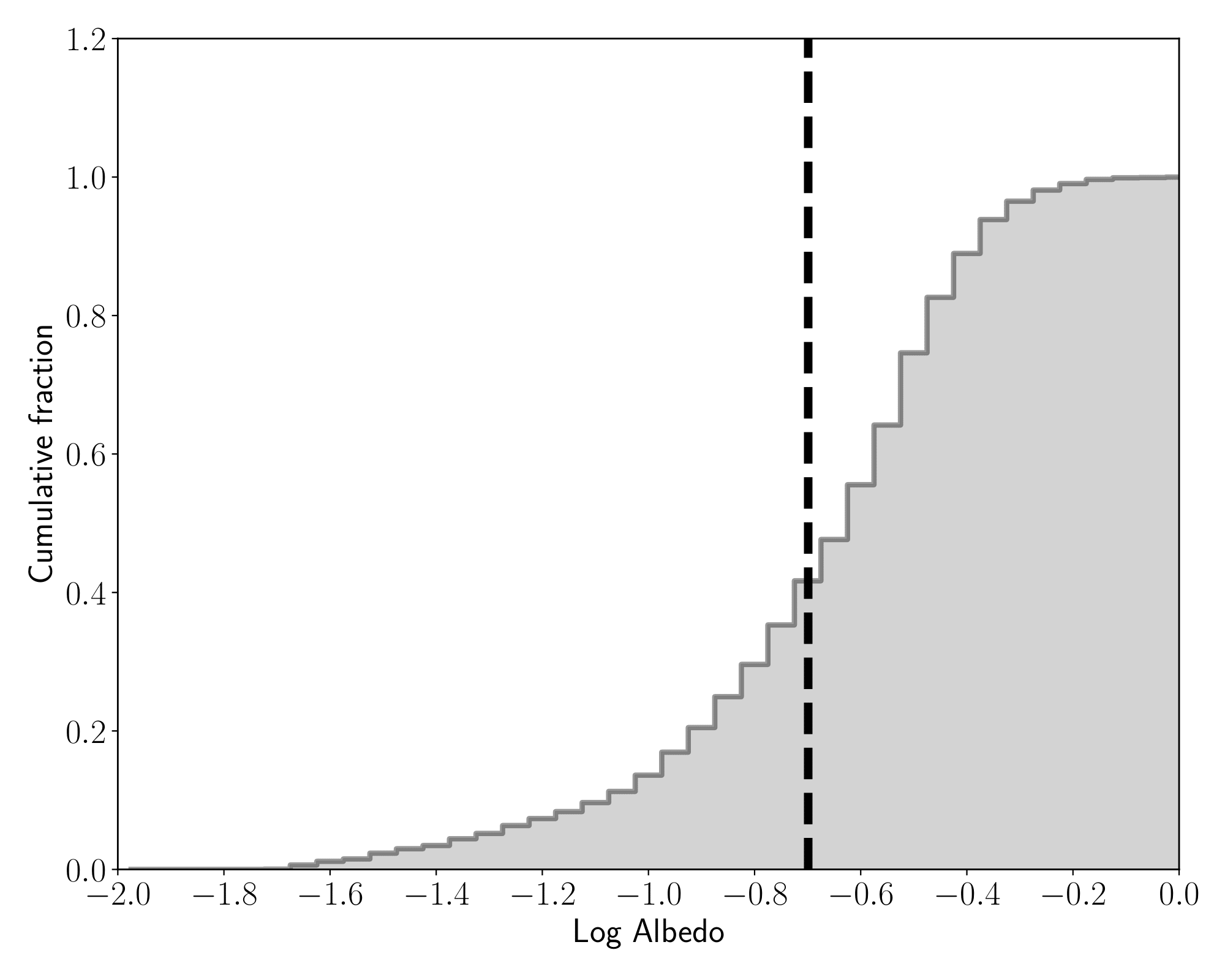}
\caption{The cumulative distribution of inner Venus objects with 15$<$$H$$<$18 computed from the NEA albedo model. The predicted albedo of \an is indicated by the dashed vertical line.}
\end{figure}


\bsp	
\label{lastpage}
\end{document}